\documentclass[aps,prb,groupedaddress,twocolumn,showpacs]{revtex4}
\usepackage{graphicx,amsmath,amssymb,bm,hyperref,color,engord,times}
\usepackage{epstopdf}
\definecolor{blue}{rgb}{0.0,0.0,1.0}
\definecolor{black}{rgb}{0.0,0.0,0.0}
\definecolor{red}{rgb}{1.0,0.0,0.0}
\hypersetup{colorlinks=true,breaklinks, linkcolor=blue,urlcolor=blue,citecolor=red}
\begin{document}
\title{Tunneling into low-dimensional and strongly correlated conductors:\\ A saddle-point approach}
\author{Kelly R. Patton}
\email[\hspace{-1.4mm}]{kpatton@physnet.uni-hamburg.de}
\affiliation{I. Institut f\"ur Theoretische Physik Universit\"at Hamburg, Hamburg 20355, Germany }
\date{\today}
\begin{abstract}
A general nonperturbative theory of the low-energy electron propagator is developed and used to calculate the single-particle density of states in a variety of systems.  This method involves the decoupling of the electron-electron interaction through a Hubbard-Stratonovich transformation, followed by a saddle-point approximation of the remaining functional integral. The final expression is found to be the tunneling analog of the infrared catastrophe that occurs in the x-ray edge problem; here, the host system responds to the potential produced by the abrupt addition of an electron during a tunneling event.  This response can lead to a suppression in the tunneling density of states near the Fermi energy.  This method is adaptable to lattice or continuum models of any dimensionality, with or without translational invariance.  When applied, the exact density of states is obtained for the Tomonaga-Luttinger model, and the pseudogap of a fractional quantum Hall fluid is recovered.            
\end{abstract}
\maketitle

\section{Introduction}
In condensed matter, tunneling experiments provide a rich source of information.  From the atomic scale analysis of a scanning tunneling microscope to the  tunneling behavior of a macroscopic supercurrent, the tunneling behavior probes intrinsic  properties of the sample and  provides opportunities to test theory against experiment, as well as generate new questions.  In the case of local tunneling, the tunneling current, or more specifically the conductance, is governed by the local electronic density of states (DOS) of the system of interest; thus,  DOS measurements by tunneling experiments have become a common and indispensable tool of condensed matter physics. Therefore, it is of great importance to be able to theoretically describe the tunneling behavior of such systems.   To make this connection from the  theoretical side, one has to calculate the relevant electron propagator of the system;  as the DOS is related to a single-particle Green's function.   
  
Besides standard diagrammatic techniques, there are many analytical and numerical  methods  available to calculate the DOS,  such as density functional methods,\cite{DFTBookgross} which can reliably  provide the DOS over a wide-energy range for weakly or non-correlated Fermi liquids.  When correlations or electron-electron interactions become strong other techniques exist, such as bosonization\cite{HaldaneJPhysC81b} and  density matrix renormalization group\cite{WhitePRL92} (DMRG) for one-dimesional systems and more recently dynamic mean field theory\cite{GeorgesRMP96} (DMFT) and its extensions\cite{RubtsovPRB08} in higher dimensions.  Although, there have been tremendous strides in the area of strongly correlated electrons, there are physical systems where all current methods breakdown.   Then one is usually  left only with exact diagonalization, which by its very nature is severely restricted  by the exponentially increasing  size of the Hilbert space.   One such system is the edge of a quantum Hall bar,\cite{GoldmanPRL01,WanPRL02,WanPRL05} where the experiments of  Grayson {\it et al.},\cite{GraysonPRL98} on the tunneling spectra at the edge of a sharply confined two-dimensional quantum Hall system, are still theoretically unresolved.  Thus, it is important to develop new methods and new insights into the tunneling behavior of such systems.  In the follow sections we outline and demonstrate a novel approach to accurately calculate the low-energy DOS in a wide variety of cases, from Fermi and Luttinger liquids to the bulk fractional quantum Hall system.  Ultimately it is hoped to apply the method developed here to the quantum Hall edge, where existing theoretical methods fail. 

In this article we derive an expression for a general fully interacting single-particle Green's function  by performing a Hubbard-Stratonovich transformation, decoupling the electron-electron interaction, followed by a saddle-point approximation of the resulting functional integral.  
The final expression provides a physically appealing picture that brings together semi-classical charge spreading theories\cite{LevitovJETP97} and x-ray edge physics,\cite{MahanPR67,NozieresPR69,AndersonPRL67,HamannPRL71,YuvalPRB70} as related to tunneling into low-dimensional and or strongly correlated systems. 
Although,  an exact expression for the Green's function, within the saddle-point approximation, is obtained, an exact evaluation  is generally  not possible.  The main obstacle to this is determining the saddle-point itself, followed by evaluating  the needed electron correlation function for the saddle-point field.  Thus, we solve for  an approximate saddle-point and then proceed to evaluate  the correlation functions in the presence of this approximate field:  either exactly or by cumulant resummation of a perturbation series.   Within these approximations we find both qualitative and quantitative agreement when compared to   known results and other methods. It is also shown in Secs.~\ref{cumulant section} and \ref{x-ray edge section} that this approach absorbs our previous, physically motived but somewhat ad hoc, approaches\cite{PattonJPhysCon08a,PattonPRB06b,PattonPRB06c,PattonPRB05} to this line of investigation, by deriving them as limiting cases of the method presented here.  Finally, in Sec.~\ref{application to LLL} we apply the full formalism to calculate the tunneling DOS  of a bulk fractional quantum Hall system,  where we  recover the experimentally observed pseudogap in the DOS.

\section{saddle-point approximation}
\label{saddle point approximation}
We start by considering a general $D$-dimensional interacting electron system, possibly in an 
external magnetic field; the second quantized  grand-canonical Hamiltonian is
\begin{align}
&H = \sum_\sigma \int d^{D}r\,  \Psi^\dagger_\sigma({\bf r}) 
 \bigg[{\Pi^2 \over 2m} + v_0({\bf r}) - \mu_0 \bigg] \Psi^{}_\sigma
({\bf r}) \nonumber \\& +\frac{1}{2}\sum_{\sigma \sigma'} \int 
d^{D}r d^{ D}r' \, \Psi_\sigma^\dagger({\bf r}) \Psi_{\sigma'}^\dagger({\bf r}') U({\bf r}-{\bf r}') 
\Psi^{}_{\sigma'}({\bf r}') \Psi_{\sigma}({\bf r}), 
\label{hamiltonian}
\end{align}
where ${\bf \Pi} = {\bf p} + {\textstyle{e \over c}} {\bf A}$, and
$v_0({\bf r})$ is any single-particle potential, which may include a
periodic lattice potential or disorder or both.  Here, $e$ is the magnitude of the electron charge, i.e.,\ $e=|e|$. Apart from an additive 
constant we can rewrite $H$ as $H_0 + V$, where
\begin{equation}
H_0 =\sum_\sigma \int d^{\scriptscriptstyle D}r \ \Psi^\dagger_\sigma(
{\bf r}) \, \big[{\textstyle {\Pi^2 \over 2m}} + v({\bf r}) - \mu \big] 
\Psi^{}_\sigma ({\bf r})
\end{equation}
and
\begin{equation} 
\label{hartree interaction}
V = {\textstyle{1 \over 2}} \int d^{\scriptscriptstyle D}r \ 
d^{\scriptscriptstyle D} r' \, \delta n({\bf r}) \, U({\bf r}-{\bf r}') \, 
\delta n({\bf r}').
\end{equation}
$H_0$ is then the Hamiltonian in the Hartree approximation. The single-particle 
potential $v({\bf r})$ includes the Hartree interaction with a 
self-consistent density $n^{}_0({\bf r})$,
\begin{equation}
v({\bf r}) = v^{}_0({\bf r}) + \int d^{\scriptscriptstyle D}r' \, U({\bf r}-
{\bf r}') \, n^{}_0({\bf r}'), 
\end{equation}
with
\begin{equation}
n^{}_0({\bf r}) =\sum_\sigma \big\langle   \Psi^\dagger_\sigma({\bf r}) 
\Psi_\sigma({\bf r}) \big\rangle_0,
\end{equation}
where the chemical potential in $H_0$ has been shifted by $-U(0)/2$, and 
$\langle O \rangle^{}_0 = {\rm Tr}  (e^{-\beta H_0} O) / {\rm Tr} (e^{-\beta
H_0})$ denotes an expectation value with respect to the Hartree-level 
Hamiltonian.  In a translationally invariant system the equilibrium density is
unaffected by interactions, but in a disordered or inhomogeneous system it 
will be necessary to distinguish between the approximate Hartree and the exact
density distributions. The interaction in (\ref{hartree interaction}) is written in
terms of density fluctuations;
\begin{equation}
\delta n({\bf r}) = \sum_\sigma \Psi^\dagger_\sigma({\bf r}) 
\Psi_\sigma({\bf r}) \ - \ n_0({\bf r}).
\end{equation}

We want to calculate the time-ordered Euclidean propagator ($\hbar=1$)
\begin{equation}
G({\bf r}_{\rm f}\sigma_{\rm f},{\bf r}_{\rm i} \sigma_{\rm i},\tau_0) =
 - \big\langle T \Psi_{\sigma_{\rm f}}({\bf r}_{\rm f} 
, \tau_0) {\overline \Psi}_{\sigma_{\rm i}}({\bf r}_{\rm i} ,0) \big\rangle_{H}
\label{G definition}
\end{equation}
for large $\tau_{0}$ or equivalently at low energy. From which, the local DOS 
\begin{equation}
N({\bf r},\omega)=-\frac{1}{\pi}\sum_{\sigma}{\rm Im}\, G({\bf r}\sigma,{\bf r}\sigma,i\omega\rightarrow\omega+i0^+)
\end{equation}
is obtained by analytic continuation of the Fourier transform of the local Green's function.
 In the interaction representation with respect to $H_0$,
\begin{equation}
G({\bf r}_{\rm f} \sigma_{\rm f}, {\bf r}_{\rm i} \sigma_{\rm i},\tau_0) = - 
{\big\langle T \Psi^{}_{\sigma_{\rm f}}({\bf r}_{\rm f}  ,\tau_0) {\overline \Psi}^{}_{ \sigma_{\rm i}}
({\bf r}_{\rm i} ,0) e^{- \int_0^\beta d \tau \, V(\tau)} 
\big\rangle_0 \over \big\langle T e^{- \int_0^\beta d \tau \, V(\tau)} 
\big\rangle_0 }.
\label{interaction representation} 
\end{equation}
Next we introduce a Hubbard-Stratonovich transformation of the form\footnote{This follows from the identity
$$\int dx_1 \dots dx_N \, e^{-{1 \over 2} x_i A_{ij} x_j} e^{x_i J_i}
= (2 \pi)^{N \over 2} \, ({\rm Det} \, A)^{-{1 \over 2}} \, 
e^{{1 \over 2} J_i A_{ij}^{-1} J_j},$$
where $A$ is a symmetric matrix.}
\begin{align}
e^{-{1 \over 2} \int \delta n \,  U \, \delta n} =\frac{\displaystyle{\int D\phi \ 
e^{-{1 \over 2} \int \phi U^{-1} \phi} \, e^{i\int \phi \, \delta n}}}{\displaystyle{ 
\int D\phi \ e^{-{1 \over 2} \int \phi  U^{-1} \phi}}},
\label{Hubbard-Stratonovich transformation}
\end{align}
where the auxiliary fields are real-scalar functions, i.e., $\phi:\mathbb{R}^{D+1}\rightarrow\mathbb{R}$. To make the transformation well-defined the fields should satisfy 
\begin{equation}
\label{condtion}
\int d^{D}r d\tau d^{D}r' d\tau'\, \phi({\bf r},\tau)U^{-1}({\bf r}\tau,{\bf r}\tau')\phi({\bf r}',\tau')>-\infty.
\end{equation}
Here, the inverse denotes a functional inverse; 
$\int d\xi\, f(x,\xi)f^{-1}(\xi,y)=\delta(x-y)$ and $U({\bf r},\tau)=U({\bf r})\delta(\tau)$.
This Hubbard-Stratonovich transformation can be understood, from a quantum electrodynamics (QED) point of view, as a reintroduction of photons as the mediator of the electron-electron interaction in the $c\rightarrow\infty$ limit. 
Using (\ref{Hubbard-Stratonovich transformation}) in (\ref{interaction representation}) gives
\begin{equation}
G({\bf r}_{\rm f} \sigma_{\rm f}, {\bf r}_{\rm i} \sigma_{\rm i},\tau_0) = 
{\cal N} \,\frac{\displaystyle{ \int D\phi \ e^{-{1 \over 2} \int \phi U^{-1} \phi} \
g({\bf r}_{\rm f} \sigma_{\rm f}, {\bf r}_{\rm i} \sigma_{\rm i},
\tau_0 | \phi)}}{\displaystyle{ \int D\phi \ e^{-{1 \over 2} \int \phi  U^{-1} \phi}}}, 
\label{exact form}
\end{equation}
where
\begin{align}
&g({\bf r}_{\rm f} \sigma_{\rm f}, {\bf r}_{\rm i} \sigma_{\rm i},\tau_0
| \phi)=\nonumber \\ &  - \big\langle T \Psi^{}_{\sigma_{\rm f}}({\bf r}_{\rm f},\tau_0)
{\overline \Psi}^{}_{\sigma_{\rm i}}({\bf r}_{\rm i},0) \, 
e^{i \int_0^\beta d \tau \int d^{\scriptscriptstyle D}r \, \phi({\bf r},\tau)
\, \delta n({\bf r},\tau) } 
\big\rangle_0 
\label{correlation function definition}
\end{align}
is a noninteracting correlation function in the presence of a purely imaginary scalar potential $i\phi({\bf r},\tau)$.  ${\cal N} = \langle T 
\exp(-\int_0^\beta d \tau \, V) \rangle_0^{-1}$ is a constant, independent of $\tau_{0}$, and will remain an unknown overall prefactor. 

Because the bare interaction $U$ in (\ref{hamiltonian}) is spin-independent, the spin dynamics remain trivial; therefore, spin labels will be suppressed, except where needed for completeness or clarity.  The inclusion of  spin related interactions within this formalism is possible. One simply needs to construct the appropriate spin-dependent Hubbard-Stratonovich transformation. 

Equation (\ref{exact form}) can be written as 
\begin{equation}
\label{action function for g}
G({\bf r}_{\rm f},{\bf r}_{\rm i},\tau_{0})={\cal N}\frac{\displaystyle{\int} D\phi\, e^{-{S}[\phi]}}{\displaystyle{\int} D\phi \ e^{-{1 \over 2} \int \phi  U^{-1} \phi}},
\end{equation}
with an action 
\begin{align}
\label{phi action}
&{S}[\phi]=\nonumber\\&\frac{1}{2}\int d\tau d\tau' d^{D}r d^{D}r'\phi({\bf r},\tau)U^{-1}({\bf r}-{\bf r}',\tau-\tau')\phi({\bf r}',\tau')\nonumber\\&-\ln g({\bf r}_{\rm f},{\bf r}_{\rm i},\tau_{0}|\phi).
\end{align}
Although, mathematically unjustified in the absence of a large or small parameter
we evaluate the functional integral by a saddle-point approximation.  We define the  saddle-point $\phi_{\rm sp}$ as the field where the action is stationary, i.e., $\frac{\delta {S}}{\delta \phi}\big|_{\phi_{\rm sp}}=0$.  This requires the saddle-point field to satisfy 
\begin{widetext}
\begin{equation}
\phi_{\rm sp}({\bf r},\tau)=\frac{-i}{g({\bf r}_{\rm f},{\bf r}_{\rm i},\tau_{0}|\phi_{\rm sp})}\int d^{D}r'\, U({\bf r}-{\bf r}')\big<T\Psi({\bf r}_{\rm f},\tau_{0})\overline{\Psi}^{}({\bf r}_{\rm i},0)\delta n({\bf r}',\tau)e^{i\int d\tau d^{D}r\, \phi_{\rm sp}({\bf r},\tau)\delta n({\bf r},\tau)}\big>_{H_{0}}.
\label{spa field}
\end{equation}
Expanding the right side  in powers of $\phi_{\rm sp}$ gives,
\begin{equation}
\label{saddle point eq}
\phi_{\rm sp}({\bf r},\tau)=\int d^{D}r'\, U({\bf r}-{\bf r}')\left[C_{1}({\bf r}',\tau)+2\int d^{D}r''d\tau''\,C_{2}({\bf r}'\tau,{\bf r}''\tau'')\phi_{\rm sp}({\bf r}'',\tau'')+\cdots\right],
\end{equation}
where $C_{1}$ and $C_{2}$ are the first and second cumulants of a cumulant expansion of (\ref{correlation function definition}) defined by
\begin{equation}
g({\bf r}_{\rm f},{\bf r}_{\rm i},\tau_{0}|\phi)=G_{0}({\bf r}_{\rm f},{\bf r}_{\rm i},\tau_{0})\exp\left[{\int C_{1}({\bf r}\tau)\phi({\bf r},\tau)+\iint\, C_{2}({\bf r}\tau,{\bf r}'\tau')\phi({\bf r},\tau)\phi({\bf r}',\tau')+\cdots}\right],
\end{equation} 
and are completely giving in terms of noninteracting Green's functions $G_{0}({\bf r},\tau)$; 
\begin{equation}
C_{1}({\bf r},\tau)=-i\frac{G_{0}({\bf r}_{\rm f},{\bf r},\tau_{0}-\tau){G_{0}(\bf r},{\bf r}_{\rm i},\tau)}{G_{0}({\bf r}_{\rm f},{\bf r}_{\rm i},\tau_{0})}
\end{equation}
and
\begin{align}
\label{c2}
C_{2}({\bf r}\tau,{\bf r}'\tau')=\frac{1}{2}\chi({\bf r},{\bf r}',\tau,\tau')-\frac{1}{2}C_{1}({\bf r},\tau)C_{1}({\bf r}',\tau')-\frac{1}{2G_{0}({\bf r}_{\rm f},{\bf r}_{\rm i},\tau_{0})}&\Big[G_{0}({\bf r},{\bf r}',\tau-\tau')G_{0}({\bf r}',{\bf r}_{\rm i},\tau')G_{0}({\bf r}_{\rm f},{\bf r},\tau_{0}-\tau)\nonumber \\ &+G_{0}({\bf r}',{\bf r},\tau'-\tau)G_{0}({\bf r},{\bf r}_{\rm i},\tau)G_{0}({\bf r}_{\rm f},{\bf r}',\tau_{0}-\tau')\Big],
\end{align}
\end{widetext}
where $\chi({\bf r},{\bf r}',\tau,\tau')=-\left<T\delta n({\bf r},\tau)\delta n({\bf r}',\tau')\right>_{H_{0}}$ is the density-density correlation function.

The exact solution of (\ref{saddle point eq}) can formally be  written as
\begin{equation}
\label{exact saddle point}
\phi_{\rm sp}({\bf r},\tau)=i\int d^{D}r'd\tau'\, U_{\rm eff}({\bf r}\tau,{\bf r}'\tau')\rho({\bf r}',\tau'),
\end{equation}
where 
\begin{align}
\label{Ueff}
U_{\rm eff}({\bf r}\tau,{\bf r}'\tau')&=\nonumber \\ &\left[U^{-1}({\bf r}-{\bf r}',\tau-\tau')-2C_{2}({\bf r}\tau,{\bf r}'\tau')+\cdots\right]^{-1}
\end{align}
is a highly non-linear inverse operator, and
\begin{equation}
\label{rho}
\rho({\bf r},\tau)=-iC_{1}({\bf r},\tau).
\end{equation}
Expanding the action, Eq.~(\ref{phi action}), about the saddle-point field to second order  and performing the functional integral \footnote{In general $\phi_{\rm sp}$ is a complex valued function, $\phi_{\rm sp}:\mathbb{R}^{D+1}\rightarrow\mathbb{C}$, and is no longer in the domain of integration.  Technically one has to deform the functional contour to pass through the saddle-point by $\phi\rightarrow\phi+\phi_{\rm sp}$ and then expand the integrand around $\phi=0$. See Ref.~[\onlinecite{PattonPRB05}] for details.}
 gives
\begin{equation}
\label{main result}
G({\bf r}_{\rm f},{\bf r}_{\rm i},\tau_{0})={\cal A}({\bf r}_{\rm f},{\bf r}_{\rm i},\tau_{0})g({\bf r}_{\rm f},{\bf r}_{\rm i},\tau_{0}|\phi_{\rm sp}),
\end{equation}
as the saddle-point approximation for the interacting Green's function,
where 
\begin{equation}
\label{fluctuation term}
{\cal A}({\bf r}_{\rm f},{\bf r}_{\rm i},\tau_{0})={\cal N}\sqrt{\frac{\det\left[U^{-1}\right]}{\det\left[\delta^{2}{S}/\delta \phi^{2}\right]}}\,e^{-\frac{1}{2}\iint \phi_{\rm sp} U^{-1}\phi_{\rm sp}}
\end{equation}
contains the fluctuation determinant.   It is believed, but left unproven,  that this prefactor  has weak $\tau_{0}$ dependance and therefore will be neglected in any applied calculation.
Equation (\ref{main result}) can also be written as
\begin{align}
\label{exact dyson}
&G({\bf r}_{\rm f},{\bf r}_{\rm i},\tau_{0})={\cal A}(\tau_{0})Z_{\rm sp}(\tau_{0})G_{\rm sp}({\bf r}_{\rm f},{\bf r}_{\rm i},\tau_{0})e^{-i\int \phi_{\rm sp}n_{0}},
\end{align}
where $Z_{\rm sp}(\tau_{0})=\big<Te^{i\int d\tau d^{D}r\,\phi_{\rm sp}({\bf r},\tau)n({\bf r},\tau)}\big>_{H_{0}}$, and the saddle-point Green's function
\begin{align}
\label{saddle point Green's function}
&G_{\rm sp}({\bf r}_{\rm f},{\bf r}_{\rm i},\tau_{0})=\nonumber \\ &-\frac{\left<T\Psi({\bf r}_{\rm f},\tau_0)
{\overline \Psi}({\bf r}_{\rm i},0) \, 
e^{i \int_0^\beta d \tau \int d^{\scriptscriptstyle D}r \, \phi_{\rm sp}({\bf r},\tau)
\,  n({\bf r},\tau) } \right>_{H_{0}}}{Z_{\rm sp}},
\end{align}
satisfies the  Dyson equation
\begin{align}
\label{saddle point dyson equation}
&G_{\rm sp}({\bf r},{\bf r}',\tau,\tau')=G_{0}({\bf r},{\bf r}',\tau-\tau')-\nonumber \\
&i\int d\tau'' d^{D}r''\,G_{0}({\bf r},{\bf r}'',\tau-\tau'')\phi_{\rm sp}({\bf r}'',\tau'')G_{\rm sp}({\bf r}'',{\bf r}',\tau'',\tau').
\end{align}

Alternatively, using a cumulant expansion for $g({\bf r}_{\rm f},{\bf r}_{\rm i},\tau_{0}|\phi_{\rm sp})$, 
\begin{align}
\label{saddle point little g}
&g({\bf r}_{\rm f},{\bf r}_{\rm i},\tau_{0}|\phi_{\rm sp})=G_{0}({\bf r}_{\rm f},{\bf r}_{\rm i},\tau_{0})\nonumber \\&\times e^{\int C_{1}({\bf r}\tau)\phi_{\rm sp}({\bf r},\tau) +\iint\, C_{2}({\bf r}\tau,{\bf r}'\tau')\phi_{\rm sp}({\bf r},\tau)\phi_{\rm sp}({\bf r}',\tau')+\cdots},
\end{align}
along with the formally exact expression for the saddle-point, Eq.~(\ref{exact saddle point}),
the   interacting Green's function in the saddle-point approximation can also be written as
\begin{align}
\label{exact cumulant}
&G({\bf r}_{\rm f},{\bf r}_{\rm i},\tau_{0})={\cal A}(\tau_{0})G_{0}({\bf r}_{\rm f},{\bf r}_{\rm i},\tau_{0})\nonumber \\ &\times e^{-1/2\int d\tau d\tau' d^{D}r d^{D}r'\, \rho({\bf r},\tau)U_{\rm eff}({\bf r}\tau,{\bf r}'\tau')\rho({\bf r}',\tau')}.
\end{align}
Equation (\ref{exact dyson}) or equivalently (\ref{exact cumulant}) is the main result of this work and the starting point for the following sections. 

Although, formally exact, within the saddle-point approximation, an exact evaluation of (\ref{exact dyson}) or (\ref{exact cumulant}) is beyond reach, as it involves finding the solution to a non-linear integral equation of infinite order for the saddle-point field, Eq.~(\ref{saddle point eq}).  Even so, making the simplest of approximations for $U_{\rm eff}$, or equivalently $\phi_{\rm sp}$, captures a great deal of nontrivial physics.  In the following  sections we  analyze  both expressions [Eq.~(\ref{exact dyson}) and (\ref{exact cumulant})]  under various approximations  and explore the underlying  physics of the results.

\section{Cumulant approximation of saddle-point equations and charge spreading}
\label{cumulant section}
\begin{figure}
\includegraphics[width=.5\textwidth]{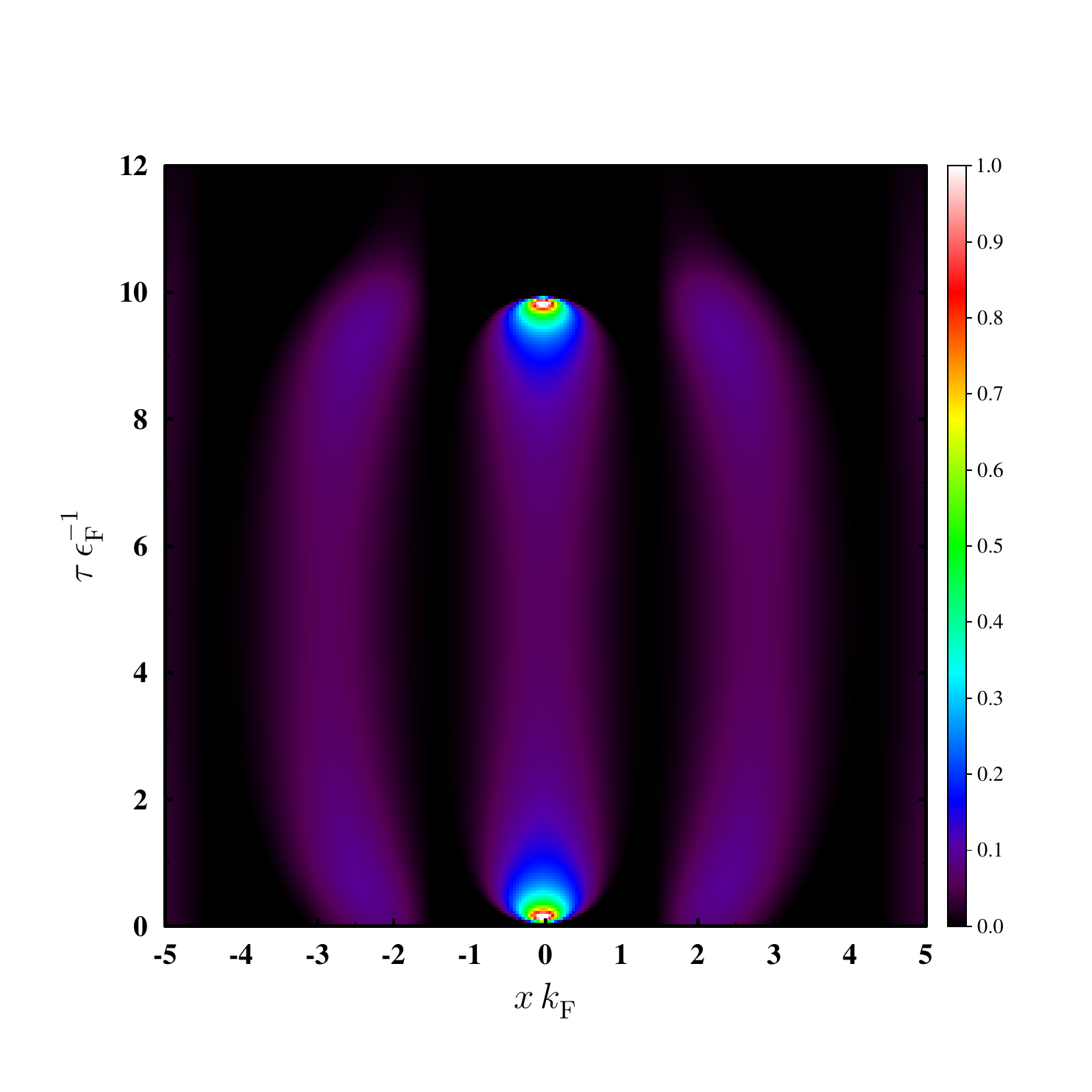}%
\vspace{-.5cm}
\caption{(Color online)  Normalized space-time contour plot of the charge density, Eq.~(\ref{rho}), for a 1-D electron gas with ${\bf r}_{\rm f}={\bf r}_{\rm i}=0$ and $\tau^{}_{0}\epsilon^{-1}_{\rm F}=10$.  Notice, at the points (${\bf r}_{\rm i},0$) and  (${\bf r}_{\rm f},\tau_{0}$) it is highly localized, in accordance with the addition and removal of an electron.  The stripes or bands are caused by Friedel-like oscillations.  See Ref.~[\onlinecite{PattonPRB06b}] for further details.    \label{fig1}}
\end{figure}

From a pure aesthetic  point of view the cumulant representation, Eq.~(\ref{exact cumulant}), seems to have the most physically appealing form.  Within the suggestive notation of (\ref{exact cumulant}), it implies the effect of interactions is to renormilize the noninteracting Green's function by an action term $e^{-S_{\rm c}(\tau_{0})}$, where 
\begin{equation}
S_{\rm c}=\frac{1}{2}\int\limits_{0}^{\beta} d\tau d\tau' \int d^{D}r d^{D}r'\, \rho({\bf r},\tau)U_{\rm eff}({\bf r}\tau,{\bf r}'\tau')\rho({\bf r}',\tau'),
\end{equation}
that describes a space and time dependent charge density $\rho$ interacting through   $U_{\rm eff}$.  In fact, in Ref.~[\onlinecite{PattonPRB06b}] a very similar expression was found, although arrived at from an almost entirely  different approach.  There it was shown that $\rho$ behaves similar to what one could loosely interpret as the charge density of the tunneling electron, or more precisely the added and removed electron associated with the Green's function (\ref{G definition}).  In other words, $\rho$ is highly localized around the space-time points (${\bf r}_{\rm i},0$) and  (${\bf r}_{\rm f},\tau_{0}$) and in between  it de-localizes,  as it relaxes and interacts with the system. 
As an example, Fig.~\ref{fig1} shows a space-time contour plot of this ``charge density'' for a one-dimenstional electron gas. 

Returning to Eq.~(\ref{exact cumulant}), as has been mentioned and can be seen, finding the exact effective interaction $U_{\rm eff}$ is unobtainable. So one is force to make approximations.  The simplest of which is replacing the effective interaction by the bare interaction $U_{\rm eff}\rightarrow U$,  this neglects the higher cumulants and other terms of (\ref{Ueff}) [or (\ref{spa field})].  This actually involves two separate approximations for Eq.~(\ref{exact cumulant}), one for the saddle-point and the other for $g({\bf r}_{\rm f},{\bf r}_{\rm i},\tau_{0}|\phi_{\rm sp})$.  This is because to obtain  (\ref{exact cumulant}) we assumed knowledge of the exact solution for the saddle-point field.  Upon making an approximation for the saddle-point the term $e^{-S_{\rm c}}$ becomes a selective resummation of perturbation theory for  $g({\bf r}_{\rm f},{\bf r}_{\rm i},\tau_{0}|\phi_{\rm sp})$.   Ultimately, this is what was effectively done in Ref.~[\onlinecite{PattonPRB06b}]  and the failures of the formalism presented there, specifically in the quantum Hall system, can be traced back to these {\it two} approximations.  Nevertheless, it was also shown that even within these relatively simple approximations the formalism was powerful enough to recover Fermi liquid theory in 2- and 3-D systems, while also obtaining the {\it exact} DOS of the 1-D Tomonaga-Luttinger model. 

If needed, as in Sec.~\ref{application to LLL}, to go beyond this two pronged approximation one should solve for the full $g({\bf r}_{\rm f},{\bf r}_{\rm i},\tau_{0}|\phi_{\rm sp})$, for an approximate $\phi_{\rm sp}$.  Doing this and tunneling's connection to the infrared catastrophe is developed next.   

\section{Connection to the x-ray edge problem and the x-ray edge limit}
\label{x-ray edge section}
Here, it is shown that the saddle-point field is a generalized x-ray  edge potential.  An x-ray edge potential is a localized time dependent field; generally  of the form 
\begin{equation}
\label{xr edge field}
\phi_{\rm xr}({\bf r},\tau)=U({\bf r})\Theta(\tau_{0}-\tau)\Theta(\tau).
\end{equation}
An infrared catastrophe can be caused by the singular screening response of a system to potentials of the form of (\ref{xr edge field}). This behavior is known to be responsible for the singular x-ray optical and photoemission spectra of metals,\cite{MahanPR67,Nozieres&DeDominicisPR69,OhtakaRMP90} Anderson's orthogonality catastrophe,\cite{AndersonPRL67,HamannPRL71} and even the Kondo effect.\cite{AndersonPRL69,YuvalPRB70} 

The x-ray edge problem, the singular absorption spectra of metals, was predicted by Mahan,\cite{MahanPR67} while studying ``excitons'' in metals; the hole of the electron-hole pair is created by the absorption of a soft x-ray.  It is assumed the hole has a large effective  mass and thus the Fermi-sea of the metal responds to  a potential of the form of  $\phi^{}_{\rm xr}$, where the hole is generated at $\tau=0$ by the absorption of a photon and then later recombines at $\tau=\tau_{0}$.

Note, if $\phi_{\rm sp}=i\phi_{\rm xr}$, apart from phase factors, Eq.~(\ref{exact dyson}) is equivalent to the x-ray edge problem,\cite{Nozieres&DeDominicisPR69} where $G_{\rm sp}$ contains the ``exitonic'' part and $Z_{\rm xp}$ is the Anderson orthogonality contribution.  This connection between tunneling and the x-ray edge field can most easily be seen by starting with the approximate (for transparency only) saddle-point field 
\begin{equation}
\phi_{\rm sp}({\bf r},\tau)=i\int d^{D}r'\, U({\bf r}-{\bf r}')\rho({\bf r}',\tau).
\end{equation}
 As was covered in the previous section, $\rho$ plays the role of an effective charge density of the tunneling electron. Expanding it in multipole moments\footnote{This can only be done in the absence of ground-state degeneracy. In such cases one has to first solve Eq.~(\ref{saddle point eq}) and then expand $\phi_{\rm sp}$ in multipole moments. For example, see Sec. \ref{application to LLL}.}
\begin{equation}
\rho({\bf r},\tau)=p^{0}(\tau)\delta({\bf r})+p^{1}(\tau)\hat{{\bf r}}{\cdot}\nabla\delta({\bf r})+\cdots,
\end{equation}
where $p^{n}(\tau)=\int d^{D}r\, r^{n}\rho({\bf r},\tau)$.
It can easily be  shown\cite{PattonPRB06b} that $p^{0}=\Theta(\tau_{0}-\tau)\Theta(\tau)$. 
By keeping only the monopole term,  the saddle-point field is then  
\begin{equation}
\label{sp equals xr}
\phi_{\rm sp}({\bf r},\tau)\approx i U({\bf r},0)\Theta(\tau_{0}-\tau)\Theta(\tau).
\end{equation}
This is the x-ray edge potential, which corresponds to the potential produced if the tunneling particle had an infinite mass. Indeed the saddle-point field is a generalized $\phi_{\rm xr}$, that accounts for both dynamic interactions and recoil of the particle.

We define the so-called x-ray edge limit as the limit where the approximate saddle-point field is taking to be of the form of (\ref{sp equals xr}).  Thus, the fully interacting Green's function in this  limit is
\begin{equation}
G({\bf r}_{\rm f},{\bf r}_{\rm i},\tau_{0})\sim g({\bf r}_{\rm f},{\bf r}_{\rm i},\tau_{0}|i\phi_{\rm xr}).
\end{equation}
or
\begin{equation}
G({\bf r}_{\rm f},{\bf r}_{\rm i},\tau_{0})\sim Z_{\rm sp}(\tau_{0})G_{\rm sp}({\bf r}_{\rm f},{\bf r}_{\rm i},\tau_{0})e^{\int \phi_{\rm xr}n_{0}}.
\end{equation}
In this limit, apart from energy shifts,  tunneling exactly reduces  to the x-ray edge problem. This  x-ray edge limit was studied in detail in Refs.~[\onlinecite{PattonPRB05,PattonPRB06c,PattonJPhysCon08a}] for a 1-D metal and the 2-D quantum Hall system, where qualitative results were found.  Of course it is desirable to go beyond this limit and include the full dynamics of the saddle-point (within a given approximation of the integral equation for $\phi_{\rm sp}$).  This can be difficult even in the simplest systems; one commonly has to find  numerical solutions to the many integral equations, if possible.

\section{Application to a fractional quantum hall fluid}
\label{application to LLL}
Because the kinetic energy, and hence the relaxation and recoil of a newly added electron, in the lowest Landau level (LLL) of a quantum Hall system is completely quenched,  
one would expect this system to be the prototypical example of an infrared catastrophe occurring during tunneling. In the LLL it is known experimentally and theoretically\cite{EisensteinPRL92,YangPRL93,HatsugaiPRL71,HePRL93,JohanssonPRL93,KimPRB94,AleinerPRL95,HaussmannPRB96,WangPRL99}  a pseudogap develops in the DOS near the Fermi energy.   In Ref.~[\onlinecite{PattonJPhysCon08a}], it was shown in the x-ray edge limit, $\phi_{\rm sp}\rightarrow\phi_{\rm xr}$, a pseudogap is indeed recovered, but this limit was introduced by hand, based only on plausible physical arguments.   It is not obvious using the full formalism presented here that such a result can be obtained.

 Here, we apply the method of the previous sections to a two-dimensional electron gas in a quantizing magnetic field.    
Choosing ${\bf B}=-B{\bf e}_{z}$, and the symmetric gauge ${\bf A}=(By/2){\bf e}_{x}-(Bx/2){\bf e}_{y}$, in the zero-temperature limit the noninteracting Green's function for the  LLL is 
\begin{equation}
G_{0}({\bf r},{\bf r}',\tau)=\Gamma({\bf r},{\bf r}')[\nu-\Theta(\tau)],
\end{equation}
where 
\begin{align}
\Gamma({\bf r},{\bf r}')&=\sum_{m}\phi^{}_{m}({\bf r})\phi^{*}_{m}({\bf r}')\nonumber\\&=\frac{1}{2\pi}\sum_{m=0}^{\infty}\frac{(rr')^{m}}{2^{m}m!}e^{im(\theta-\theta')}e^{-r^{2}/4}e^{-r'^{2}/4},
\end{align}
and $\phi^{}_{m}({\bf r})$ are the single-particle eigenfunctions (in units where the magnetic length $\ell=\sqrt{\hbar c/eB}=1$), $0<\nu<1$ is the filling factor, and $\Theta(\tau)$ is the Heaviside step function. 
The charge density, Eq.~(\ref{rho}), for this system is therefore 
\begin{equation}
\rho({\bf r},\tau)=\frac{e^{-|{\bf r}|^{2}/2}}{2\pi}[\Theta(\tau_{0}-\tau)\Theta(\tau)-\nu].
\end{equation}
In systems such as this, with ground-state degeneracy, the  leading order approximation for $\phi_{\rm sp}$, $\phi_{\rm sp}({\bf r},\tau)\approx i\int d^{D}r'\, U({\bf r}-{\bf r}')\rho({\bf r}',\tau)$, fails the integrable condition, Eq.~(\ref{condtion}), for any meaningful interaction in the $\beta\rightarrow\infty$ limit.  Therefore, one has to find another approximate solution of (\ref{saddle point eq}) by including higher-order terms, in $\phi_{\rm sp}$, to obtain a suitable  saddle-point.  At a minimum one has to solve the linearized version of (\ref{saddle point eq}).  This is what we know turn to.  Additionally, we will assume a finite inverse-temperature $\beta$, only taking $\beta\rightarrow\infty$ at the end if desired.  At finite temperature the LLL Green's function is 
\begin{equation}
G_{0}({\bf r},{\bf r}',\tau)=\Gamma({\bf r},{\bf r}')[n^{}_{\rm F}(\epsilon^{}_{0}-\mu)-\Theta(\tau)]e^{-(\epsilon^{}_{0}-\mu)\tau},
\end{equation}
where $n_{\rm F}^{}(\omega)$ is the Fermi distribution, $\epsilon^{}_{0}$ is the single-particle energy of the lowest Landau level, and  the chemical potential $\mu$ is defined such that
\begin{equation}
\nu=\sum_{n=0}^{\infty}\frac{1}{\exp[\beta(\epsilon_{n}-\mu)]+1}.
\end{equation}
Note, the Green's function is aperiodic, $G_{0}(\tau-\beta)=-G_{0}(\tau)$, as it should.
The finite-temperature charge density is then 
\begin{equation}
\rho({\bf r},\tau)=\frac{e^{-|{\bf r}|^{2}/2}}{2\pi}[\Theta(\tau_{0}-\tau)\Theta(\tau)-n_{\rm F}^{}(\epsilon^{}_{0}-\mu)].
\end{equation}

From (\ref{saddle point eq}), the linearized integral equation for $\phi_{\rm sp}$, setting ${\bf r}_{\rm f}={\bf r}_{\rm i}=0$, is 

\begin{widetext}
\begin{align}
&\phi_{\rm sp}({\bf r},\tau)=i\int d^{2}r'\,U({\bf r}-{\bf r}')\rho({\bf r'},\tau)+\int d^{2}r'd^{2}r''\int\limits_{0}^{\beta}d\tau'\,U({\bf r}-{\bf r}')\chi({\bf r}',{\bf r}'',\tau,\tau')\phi_{\rm sp}({\bf r}'',\tau')\nonumber\\&+\frac{1}{(2\pi)^{2}}\int d^{2}r'd^{2}r''\int\limits_{0}^{\beta}d\tau'\,U({\bf r}-{\bf r}')e^{-|{\bf r}'|^{2}/2}e^{-|{\bf r}''|^{2}/2}\Big\{[\Theta(\tau_{0}-\tau)\Theta(\tau)-n_{\rm F}^{}(\epsilon^{}_{0}-\mu)][\Theta(\tau_{0}-\tau')-n_{\rm F}^{}(\epsilon^{}_{0}-\mu)]\nonumber\\&-[n_{\rm F}^{}(\epsilon^{}_{0}-\mu)-\Theta(\tau-\tau')][n_{\rm F}^{}(\epsilon^{}_{0}-\mu)-\Theta(\tau_{0}-\tau)]-[n_{\rm F}^{}(\epsilon^{}_{0}-\mu)-\Theta(\tau'-\tau)][n_{\rm F}^{}(\epsilon^{}_{0}-\mu)-\Theta(\tau_{0}-\tau')]\Big\}\phi_{\rm sp}({\bf r}'',\tau'),
\end{align}
\end{widetext}
where $\chi({\bf r},{\bf r'},\tau,\tau')=-\frac{n^{\rm F}(1-n^{\rm F})}{(2\pi)^{2}}\,e^{-|{\bf r}-{\bf r}'|^{2}/2}. $
In the low-temperature limit, the solution is (see appendix \ref{quantum hall saddle-point equation}) 
\begin{equation}
\label{saddle point for strong field}
\phi_{\rm sp}({\bf r},\tau)\xrightarrow{\beta\rightarrow\infty}\frac{i}{2\pi}\int d^{2}r'\,U({\bf r}-{\bf r}')e^{-|{\bf r}'|^{2}/2}\Theta(\tau_{0}-\tau)\Theta(\tau).
\end{equation}
Because this is  a valid saddle-point field, we can now determine the interacting Green's function. 
Within the cumulant approximation, Sec.~{\ref{cumulant section}}, using (\ref{saddle point for strong field}) gives
\begin{equation}
\label{hard gap}
G(0,0,\tau_{0})\sim\frac{\nu-1}{2\pi}e^{-\lambda(1-\nu)\tau_{0}},
\end{equation}
where $\lambda=(4\pi)^{-1}\iint d^{2}rd^{2}r'U({\bf r}-{\bf r}')e^{-|{\bf r}'|^{2}/2}e^{-|{\bf r}|^{2}/2}$.  Equation (\ref{hard gap}) leads  to a delta function or  ``hard-gap'' in the DOS 
\begin{equation}
N(0,\omega)={\rm const}\times \delta[\omega-\lambda(1-\nu)],
\end{equation}
instead of the observed pseudogap.  Mathematically a pesudogap can be modeled as a function that decays faster than any power-law at low energy. To obtain this behavior implies that in the time domain the function should exhibit a similar decay rate as $\tau\rightarrow\infty$. Therefore, the fully interacting Green's function would be expected to be a nonalgebraic and or possibly nonanalytic function of $\tau_{0}$.  The cumulant approximation, which is basically a selective resummation of perturbation theory in $\phi_{\rm sp}$ or more explicitly for this case perturbation in $\lambda \tau_{0}$,  isn't powerful enough to capture such behavior; an exact solution of the Dyson equation for the saddle-point Green's function, Eq.~(\ref{saddle point dyson equation}), is required for this system.  

With (\ref{saddle point for strong field}),  the Dyson equation for the saddle-point Green's function is
\begin{widetext}
\begin{align}
&G_{\rm sp}({\bf r},{\bf r}',\tau,\tau')=\Gamma({\bf r},{\bf r}')\big[n_{\rm F}^{}(\epsilon^{}_{0}-\mu)-\Theta(\tau-\tau')\big]e^{-(\epsilon^{}_{0}-\mu)(\tau-\tau')}
\nonumber\\&+\frac{1}{2\pi}\int\limits_{0}^{\tau_{0}} d\tau_{1} d^{2}r_{1}d^{2}r_{2}\,\Gamma({\bf r},{\bf r}_{1})\big[n_{\rm F}^{}(\epsilon^{}_{0}-\mu)-\Theta(\tau-\tau_{1})\big]e^{-(\epsilon^{}_{0}-\mu)(\tau-\tau_{1})}\,U({\bf r}_{1}-{\bf r}_{2})e^{-|{\bf r}_{2}|^{2}/2}G_{\rm sp}({\bf r}_{2},{\bf r}',\tau_{1},\tau').
\end{align}
The solution of which is\cite{PattonJPhysCon08a}
\begin{equation}
G_{\rm sp}({\bf r},{\bf r}',\tau,\tau')=\sum_{m} \phi^{}_{m}({\bf r})\phi^{*}_{m}({\bf r}')a^{}_{m}(\tau,\tau'),
\end{equation}
where
\begin{equation}
a^{}_{m}(\tau,\tau')=\frac{(n_{\rm F}^{}(\epsilon^{}_{0}-\mu)-1)\Theta(\tau-\tau')+n_{\rm F}^{}(\epsilon^{}_{0}-\mu)\Theta(\tau'-\tau)\exp(-\lambda_{m}\tau_{0})}{1-n_{\rm F}^{}(\epsilon^{}_{0}-\mu)+n_{\rm F}^{}(\epsilon^{}_{0}-\mu)\exp(-\lambda_{m}\tau_{0})}\exp[-\lambda^{}_{m}(\tau-\tau')]\exp[-(\epsilon^{}_{0}-\mu)(\tau-\tau')]
\end{equation}
and
\end{widetext}
\begin{equation}
\lambda^{}_{m}=\int d^{2}rd^{2}r'\, \phi^{*}_{m}({\bf r})U({\bf r}-{\bf r}')e^{-|{\bf r}'|^{2}/2}\phi^{}_{m}({\bf r}),
\end{equation}
are the diagonal matrix elements of the saddle-point interaction in the LLL.  For a screened Coulomb potential of the form
\begin{equation}
U({\bf r})=e^{2}\frac{\exp(-\alpha|{\bf r}|)}{|{\bf r}|},
\end{equation}
\begin{equation}
\lambda^{}_{m}=e^{2}\int\limits_{0}^{\infty}dk\,\frac{k\exp(-k^{2}/2)}{\sqrt{\alpha^{2}+k^{2}}} {}^{}_{1}{F}^{}_{1}(1+m,1;-k^{2}/2),
\end{equation}
where $^{}_{1}F^{}_{1}(a,b;z)$ is the confluent hypergeometric function of the first kind.  Experimentally the screening length $\alpha$  would typically be the distance to the nearest gate or to the second 2-D electron gas in bilayer systems.  

From appendix \ref{Saddle-Point scattering matrix}
\begin{equation}
Z_{\rm sp}(\tau_{0})=\prod_{m=0}^{\infty}\big[1-n_{\rm F}^{}(\epsilon^{}_{0}-\mu)+n_{\rm F}^{}(\epsilon^{}_{0}-\mu)\exp(-\lambda_{m}\tau_{0})\big]. 
\end{equation}
This finally gives the interacting Green's function  as
\begin{align}
\label{full green's for strong field}
&G(0,0,\tau_{0})\sim Z_{\rm sp}(\tau_{0})G_{\rm sp}(0,0,\tau_{0})\exp\big( \nu \alpha^{-1}\tau_{0}\big)\nonumber\\&=\frac{1}{2\pi}\prod_{m=1}^{\infty}\big[1-n_{\rm F}^{}(\epsilon^{}_{0}-\mu)+n_{\rm F}^{}(\epsilon^{}_{0}-\mu)\exp(-\lambda_{m}\tau_{0})\big]\nonumber\\&\times(n_{\rm F}^{}(\epsilon^{}_{0}-\mu)-1)\exp(-\lambda^{}_{0}\tau_{0})\exp\big( \nu \alpha^{-1}\tau_{0}\big).
\end{align}
\begin{figure}
\includegraphics[width=.5\textwidth]{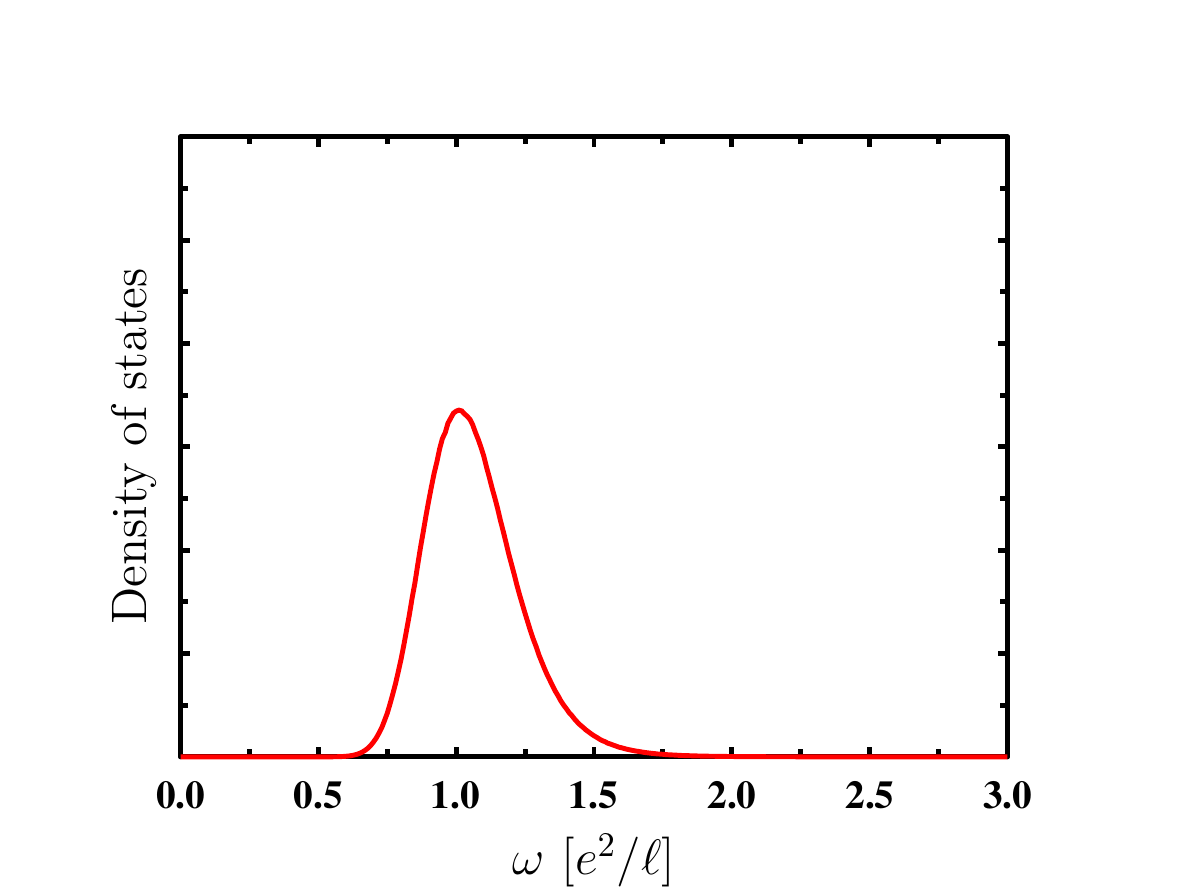}%
\caption{The DOS (up to an overall constant prefactor) of the LLL. Found by analytic continuation of (\ref{full green's for strong field}), using the experimentally relevent parameters  $\beta=10$, $\alpha =.05$, and $\nu=1/3$, in units of $e^{2}/\ell$ and $\ell$. \label{fig2}}
\end{figure}
Figure \ref{fig2} shows the DOS obtained from (\ref{full green's for strong field}), by analytically continuation using the maximum entropy method.  As can be seen, a pseudogap is present in agreement with experiment.  Although, the width of the peak is of the order seen experimentally, the actual location cannot be totally trusted. This is because we have solved for $G_{\rm sp}$ using a long time approximation for $G^{}_{0}$. Then used the solution to find $Z_{\rm sp}$, (see appendix \ref{Saddle-Point scattering matrix}) but short time information enters into  $Z_{\rm sp}$.  This is a well-known result from the x-ray edge problem, where unknown energy shifts can occur.    The inclusion of higher Landau levels in $G_{0}$ would fix this. 

\section{Conclusions}
By starting with an exact functional integral representation of an interacting Green's function, followed by a saddle-point approximation, we have shown that the low-energy tunneling behavior of electronic systems is governed by the response of the host system to the time-dependent potential produced by the newly added electron.   In the x-ray edge limit the infrared catastrophe occurring during a  tunneling event can be exactly mapped to the well-known x-ray edge problem.  Within various approximations for the saddle-point we have obtain qualitative as well as quantitative results in a wide spectrum of systems including: Fermi liquids, the Tomonaga-Luttinger liquid, and quantum Hall system.

 Future work will involve applying and comparing this method to other exactly solvable models, such as the 1-D Hubbard model and ultimately to the edge of a quantum Hall fluid, where existing theoretical methods fail. 

\begin{acknowledgments}
I would like the thank Michael Geller for many useful discussions and Hartmut Haffermann for help with the maximum entropy calculation. 
This work was supported by  the
German Research Council (DFG) under SFB 668.
\end{acknowledgments}
\appendix 

\section{Saddle-point field for the Quantum Hall system}
\label{quantum hall saddle-point equation}
Here, we solve for the saddle-point field including the linear term in $\phi_{\rm sp}$ of the integral equation (\ref{saddle point eq}).  This is needed because the simplest approximation for the saddle-point obtained by keeping only the zeroth order term of Eq.~(\ref{saddle point eq}) is not in the integration domain of the functional integral. For the quantum Hall system  this then reads
\begin{widetext} 
\begin{align}
&\phi_{\rm sp}({\bf r},\tau)=i\int d^{2}r'\,U({\bf r}-{\bf r}')\rho({\bf r'},\tau)+\int d^{2}r'd^{2}r''\int\limits_{0}^{\beta}d\tau'\,U({\bf r}-{\bf r}')\chi({\bf r}',{\bf r}'',\tau,\tau')\phi_{\rm sp}({\bf r}'',\tau')\nonumber\\&+\frac{1}{(2\pi)^{2}}\int d^{2}r'd^{2}r''\int\limits_{0}^{\beta}d\tau'\,U({\bf r}-{\bf r}')e^{-|{\bf r}'|^{2}/2}e^{-|{\bf r}''|^{2}/2}\Big\{[\Theta(\tau_{0}-\tau)\Theta(\tau)-n_{\rm F}^{}(\epsilon^{}_{0}-\mu)][\Theta(\tau_{0}-\tau')-n_{\rm F}^{}(\epsilon^{}_{0}-\mu)]\nonumber\\&-[n_{\rm F}^{}(\epsilon^{}_{0}-\mu)-\Theta(\tau-\tau')][n_{\rm F}^{}(\epsilon^{}_{0}-\mu)-\Theta(\tau_{0}-\tau)]-[n_{\rm F}^{}(\epsilon^{}_{0}-\mu)-\Theta(\tau'-\tau)][n_{\rm F}^{}(\epsilon^{}_{0}-\mu)-\Theta(\tau_{0}-\tau')]\Big\}\phi_{\rm sp}({\bf r}'',\tau').
\end{align}
Using $\chi({\bf r},{\bf r'},\tau,\tau')=-\frac{n_{\rm F}^{}(1-n_{\rm F}^{})}{(2\pi)^{2}}\,e^{-|{\bf r}-{\bf r}'|^{2}/2}$ and 
\begin{equation}
\Theta(\tau_{0}-\tau)\big[\Theta(\tau_{0}-\tau')-\Theta(\tau-\tau')\big]-\Theta(\tau'-\tau)\Theta(\tau_{0}-\tau')=0
\end{equation}
 gives
\begin{align}
\label{sd for qhe}
\phi_{\rm sp}({\bf r},\tau)&=i\int d^{2}r'\,U({\bf r}-{\bf r}')\rho({\bf r'},\tau)-\frac{n_{\rm F}^{}(\epsilon^{}_{0}-\mu)[1-n_{\rm F}^{}(\epsilon_{0}-\mu)]}{(2\pi)^{2}}\int d^{2}r'd^{2}r''\int\limits_{0}^{\beta}d\tau'\,U({\bf r}-{\bf r}')e^{-|{\bf r}'-{\bf r}''|^{2}/2}\phi_{\rm sp}({\bf r}'',\tau')\nonumber\\&+\frac{n_{\rm F}^{}(\epsilon^{}_{0}-\mu)\big[1-n_{\rm F}^{}(\epsilon^{}_{0}-\mu)\big]}{(2\pi)^{2}}\int d^{2}r'd^{2}r''\int\limits_{0}^{\beta}d\tau'\,U({\bf r}-{\bf r}')e^{-|{\bf r}'|/2}e^{-|{\bf r}''|/2}\phi_{\rm sp}({\bf r}'',\tau').
\end{align}
Introducing the Fourier transforms  
\begin{equation}
\phi_{\rm sp}({\bf k},i\Omega_{n})=\int d^{2}r\int\limits_{0}^{\beta}d\tau\,\phi_{\rm sp}({\bf r},\tau)e^{i\Omega_{n}\tau}e^{-i{\bf k}\cdot{\bf r}}\hspace{1cm}\text{and}\hspace{1cm} \phi_{\rm sp}({\bf r},\tau)=\frac{1}{\beta}\sum_{n}\int \frac{d^{2}k}{(2\pi)^{2}}\phi_{\rm sp}({\bf k},i\Omega_{n})e^{-i\Omega_{n}\tau}e^{i{\bf k}\cdot{\bf r}},
\end{equation}
where $i\Omega_{n}$ is a bonsonic frequency, the integral equation in Fourier space is 
\begin{align}
\phi_{\rm sp}({\bf k},i\Omega_{n})&=iU({\bf k})\rho({\bf k},i\Omega_{n})-\frac{n_{\rm F}^{}(\epsilon^{}_{0}-\mu)[1-n_{\rm F}^{}(\epsilon_{0}-\mu)]}{(2\pi)^{2}}U({\bf k})e^{-|{\bf k}|^{2}/2}\phi_{\rm sp}({\bf k},i\Omega_{n'})\beta\delta_{i\Omega_{n'},0}\delta_{i\Omega_{n},0}\nonumber\\&+\frac{n_{\rm F}^{}(\epsilon^{}_{0}-\mu)[1-n_{\rm F}^{}(\epsilon_{0}-\mu)]}{(2\pi)^{2}}U({\bf k})e^{-|{\bf k}|^{2}/2}\int d^{2}k\,'e^{-|{\bf k}'|^{2}/2}\phi_{\rm sp}({\bf k}',i\Omega_{n'})\beta\delta_{i\Omega_{n'},0}\delta_{i\Omega_{n},0}.
\end{align}
Thus for $i\Omega_{n}\ne0$
\begin{equation}
\phi_{\rm sp}({\bf k},i\Omega_{n}\ne0)=iU({\bf k})\rho({\bf k},i\Omega_{n}\ne0)
\end{equation}
and for $i\Omega_{n}=0$
\begin{align}
\phi_{\rm sp}({\bf k},i\Omega_{n}=0)&=iU({\bf k})\rho({\bf k},i\Omega_{n}=0)-\frac{\beta}{(2\pi)^{2}}n_{\rm F}^{}(\epsilon^{}_{0}-\mu)[1-n_{\rm F}^{}(\epsilon_{0}-\mu)] U({\bf k})e^{-|{\bf k}|^{2}/2}\phi_{\rm sp}({\bf k},i\Omega_{n}=0)\nonumber\\&+\frac{\beta}{(2\pi)^{2}}n_{\rm F}^{}(\epsilon^{}_{0}-\mu)[1-n_{\rm F}^{}(\epsilon_{0}-\mu)]U({\bf k})e^{-|{\bf k}|^{2}/2}\int d^{2}k'e^{-|{\bf k}'|^{2}/2}\phi_{\rm sp}({\bf k}',i\Omega_{n}=0).
\end{align}
or
\begin{align}
\phi_{\rm sp}({\bf k},i\Omega_{n}=0)&=\frac{iU({\bf k})\rho({\bf k},i\Omega_{n}=0)}{1+\frac{\beta}{(2\pi)^{2}}n_{\rm F}^{}(\epsilon^{}_{0}-\mu)[1-n_{\rm F}^{}(\epsilon_{0}-\mu)]  U({\bf k})e^{-|{\bf k}|^{2}/2}}\nonumber\\&+\frac{\beta}{(2\pi)^{2}}\frac{n_{\rm F}^{}(\epsilon^{}_{0}-\mu)[1-n_{\rm F}^{}(\epsilon_{0}-\mu)]  U({\bf k})e^{-|{\bf k}|^{2}/2}}{1+\frac{\beta}{(2\pi)^{2}}n_{\rm F}^{}(\epsilon^{}_{0}-\mu)[1-n_{\rm F}^{}(\epsilon_{0}-\mu)]  U({\bf k})e^{-|{\bf k}|^{2}/2}}\int d^{2}k'e^{-|{\bf k}'|^{2}/2}\phi_{\rm sp}({\bf k}',i\Omega_{n}=0)
\end{align}
This is a Fredholm integral equation with a degenerate kernel whose solution is simply given by\cite{integralequationbook}
\begin{align}
\phi_{\rm sp}({\bf k},i\Omega_{n}=0)&=\frac{iU({\bf k})\rho({\bf k},i\Omega_{n}=0)}{1+\frac{\beta}{(2\pi)^{2}}n_{\rm F}^{}(\epsilon^{}_{0}-\mu)[1-n_{\rm F}^{}(\epsilon_{0}-\mu)]  U({\bf k})e^{-|{\bf k}|^{2}/2}}\nonumber\\&+A(i\Omega_{n}=0)\frac{\beta}{(2\pi)^{2}}\frac{n_{\rm F}^{}(\epsilon^{}_{0}-\mu)[1-n_{\rm F}^{}(\epsilon_{0}-\mu)]  U({\bf k})e^{-|{\bf k}|^{2}/2}}{1+\frac{\beta}{(2\pi)^{2}}n_{\rm F}^{}(\epsilon^{}_{0}-\mu)[1-n_{\rm F}^{}(\epsilon_{0}-\mu)]  U({\bf k})e^{-|{\bf k}|^{2}/2}}
\end{align}
where 
\begin{equation}
A(i\Omega_{n}=0)=\frac{\displaystyle{\int} d^{2}k\,\frac{iU({\bf k})\rho({\bf k},i\Omega_{n}=0)e^{-|{\bf k}|^{2}/2}}{1+\frac{\beta}{(2\pi)^{2}}n_{\rm F}^{}(\epsilon^{}_{0}-\mu)[1-n_{\rm F}^{}(\epsilon_{0}-\mu)]  U({\bf k})e^{-|{\bf k}|^{2}/2}}}{\displaystyle{1-\frac{\beta}{(2\pi)^{2}}\int d^{2}k\frac{n_{\rm F}^{}(\epsilon^{}_{0}-\mu)[1-n_{\rm F}^{}(\epsilon_{0}-\mu)]  U({\bf k})e^{-|{\bf k}|^{2}/2}e^{-|{\bf k}|^{2}/2}}{1+\frac{\beta}{(2\pi)^{2}}n_{\rm F}^{}(\epsilon^{}_{0}-\mu)[1-n_{\rm F}^{}(\epsilon_{0}-\mu)]  U({\bf k})e^{-|{\bf k}|^{2}/2}}}}.
\end{equation}
Converting back to $\tau$ space 
\begin{align}
\phi_{\rm sp}({\bf k},\tau)=\frac{1}{\beta}\sum_{n\in \mathbb{Z}}iU({\bf k})\rho({\bf k},i\Omega_{n})e^{-i\Omega_{n}\tau}-\beta^{-1}iU({\bf k})\rho({\bf k},i\Omega_{n}=0)+\beta^{-1}\phi_{\rm sp}({\bf k},i\Omega_{n}=0),
\end{align} 
where  the $i\Omega=0$ term has been subtracted from the sum over frequencies. 
 In the $\beta\rightarrow\infty$ limit
\begin{equation}
\phi_{\rm sp}({\bf k},\tau)\rightarrow \frac{i}{2\pi}U({\bf k})e^{-|{\bf k}|^{2}/2}\Theta(\tau_{0}-\tau)\Theta(\tau),
\end{equation}
or in real space
\begin{equation}
\phi_{\rm sp}({\bf r},\tau)=\frac{i}{2\pi}\int d^{2}r'\,U({\bf r}-{\bf r}')e^{-|{\bf r}'|^{2}/2}\Theta(\tau_{0}-\tau)\Theta(\tau).
\end{equation}
\end{widetext}
This is the form of the saddle-point field that is used in Sec.~\ref{application to LLL}.
\section{Saddle-Point S-matrix}
\label{Saddle-Point scattering matrix}
Here, the saddle-point scattering matrix $Z_{\rm sp}$ is found from $G_{\rm sp}$.  This is a straightforward  generalization of the coupling constant integration trick found in Ref.~[\onlinecite{NozieresPR69}]. 
 
From the linked cluster theorem,
\begin{equation}
Z_{\rm sp}=\big\langle Te^{i\int d\tau d^{D}r\ \phi_{\rm sp}({\bf r},t)\hat{n}({\bf r},\tau)}\big\rangle_{0}=e^{M},
\end{equation}

where
\begin{widetext}
\begin{align}
 &M=\sum_{l=1}^{\infty} \frac{i^{l}}{l}\int d\tau_{1} d^{D}r_{1}\cdots d\tau_{l}d^{D}r_{l}\,\phi_{\rm sp}({\bf r}_{1},\tau_{1})\cdots\phi_{\rm sp}({\bf r}_{l},\tau_{l})\left<T\hat{n}({\bf r}_{1},\tau_{1})\cdots \hat{n}({\bf r}_{l},\tau_{l})\right>_{{\rm different \, connected}}
\end{align}
or
\begin{align}
 &M=i\sum_{l=1}^{\infty}\frac{1}{l}\int d\tau_{1} d^{D}r_{1}\cdots d\tau_{l}d^{D}r_{l}\, \phi_{\rm sp}({\bf r}_{1},\tau_{1})\cdots \phi_{\rm sp}({\bf r}_{l},\tau_{1})G_{0}({\bf r}_{1},{\bf r}_{l},\tau_1,\tau_l)\cdots G_0({\bf r}_{l},{\bf r}_{1},\tau_{l},\tau_{1}^+).
\end{align}
Changing the summation limit leads to
\begin{align}
\label{M}
&M=i\sum_{l=0}^{\infty}\frac{1}{l+1}\int d\tau_{1}d^{D}r_{1}\cdots d\tau_{l+1} d^{D}r_{l+1} \phi_{\rm sp}({\bf r}_{1},\tau_{1})\cdots\phi_{\rm sp}({\bf r}_{l+1},\tau_{l+1})G_{0}({\bf r}_{1},{\bf r}_{l+1},\tau_{1},\tau_{l+1})\cdots G_0({\bf r}_{l+1},{\bf r}_{1},\tau_{l+1},\tau_{1}^+).
\end{align}
Introducing 
\begin{equation}
{G}^{\xi}_{\rm sp}({\bf r},{\bf r}',\tau,\tau^+)=G_0({\bf r},{\bf r}',\tau,\tau^+)-i\,\xi \int d\tau''d^{D}r''\, G_0({\bf r},{\bf r}'',\tau,\tau'')\phi_{\rm sp}({\bf r}'',\tau'') G^{\xi}_{\rm sp}({\bf r}'',{\bf r}',\tau'',\tau^+)
\end{equation}
\end{widetext}
and using
\begin{equation}
\frac{1}{l+1}=\int_{0}^{1}d\xi \ \xi^l
\end{equation}
Eq.\ (\ref{M}) can be written simply as
\begin{equation}
M=i\int d\tau\int d^{D}r\int_0^1d\xi \, \phi_{\rm sp}({\bf r},\tau)  G^{\xi}_{\rm sp}({\bf r},{\bf r},\tau,\tau^+).
\end{equation}
Reinstating spin for completeness gives
\begin{equation}
M=i\sum_{\sigma}\int d\tau\int d^{D}r\int_0^1d\xi \, \phi_{\rm sp}({\bf r},\tau) G^{\xi}_{\rm sp}({\bf r},{\bf r},\tau,\tau^+,\sigma).
\end{equation}

\bibliography{/Users/kpatton/Bibliographies/Master,/Users/kpatton/Bibliographies/Mikes_Bibs/MRGhall,/Users/kpatton/Bibliographies/Mikes_Bibs/MRGmanybody,/Users/kpatton/Bibliographies/Mikes_Bibs/MRGbooks}

\end{document}